# On the Solar EUV Deposition in the Inner Comae of Comets with Large Gas Production Rates

Anil Bhardwaj

Space Physics Laboratory, Vikram Sarabhai Space Centre, Trivandrum 695022, India

Tel : +91-471-2562330; Fax : +91-471-2706535; email: bhardwaj_spl@yahoo.com

Current email: Anil_Bhardwaj@vssc.gov.in

abstract
**Abstract**

In this letter we have made a comparative study of degradation of solar EUV radiation and EUV-generated photoelectrons in the inner comae of comets having different gas production rates, Q, with values $1\times10^{28}$, $7\times10^{29}$, $1\times10^{31}$, and $1\times10^{32}$ s$^{-1}$. We found that in higher-Q comets the radial profile of $H_2O^+$ photo-production rate depicts a double-peak structure and that the differences in sunward and anti-sunward photoionization rates are pronounced. We show that photoelectron impact ionization is an order of magnitude larger than photoionization rate near the lower photoionization peak in comets with Q >~$1\times10^{31}$ s$^{-1}$. The present study reveals the importance of photoelectrons relative to solar EUV as the ionization source in the inner coma of high-Q comets.

## 1. Introduction

Comets are water-dominated tiny objects in our solar system. A comet exists as a bare nucleus at large heliocentric distances, but develop a huge coma and long tail when closer to the Sun (typically ~< 3 AU). The coma, which is generated by sublimation of volatile species from the nucleus, hosts most of the physical, chemical and dynamical processes for comets. One of the important parameters that govern the activity of the comet, and thus size of the coma, is the total gas production rate, Q (s$^{-1}$), which depends on the intrinsic properties of the comet, for example its size, and is a function of the heliocentric distance of the comet.

Comet C/1995 O1 Hale Bopp was by far the brightest comet of the last century. The very high gas production rate of Hale Bopp imposes several interesting physical effects that are generally presumed unimportant in lower-Q comets and are often neglected in the analysis of observations, interpretation, and modeling of data. The energetics and chemistry in the coma are initiated by the absorption of solar radiation by neutral constituents of the coma. Solar EUV radiation is the major source of ionization in the cometary coma; other sources include EUV-generated photoelectrons and auroral electrons of solar wind origin [e.g., *Boice et al.*, 1986; Cravens *et al.*, 1987; *Korosmezey et al.*, 1987; *Wegmann et al.*, 1987, 1999; Bhardwaj *et al.*, 1990, 1996; *Gan and Cravens*, 1990; *Bhardwaj*, 1999]. Photoionization of cometary molecules generates ions and photoelectrons, which initiate the ion-neutral chemistry in the coma. The radial profile of the ionization rate is therefore an important input in the coma chemistry models.

In this paper we have demonstrated that the degradation of solar EUV photons in the inner cometary coma must be properly treated in case of comets with large Q since it affects the shape of radial profiles of ion production rates, and thus will have major implications for those quantities that are related to it. Specifically, we have studied the effects of high Q-comets on radial profiles of (i) photoionization rates of $H_2O^+$ and its variation with solar zenith angle (SZA), and (ii) photoelectron impact ionization of $H_2O$. In particular, we found that in the case of high-Q comets the altitude profile of photoionization rate has a double-peak structure and that the SZA-effect has a marked bearing on the ion production rate profiles in the anti-sunward direction. We have also shown that in the inner coma of high-Q comets the photoelectron impact give rise to an additional ionization peak whose magnitude is a factor of 10 or more larger than the photoionization peak. The present results will be useful in interpretation and understanding the data in higher production rate comets, like Hale Bopp, and other bright comets. The findings will have implications for the energetics and chemistry in the inner coma of high-Q comets.

## 2. Model Calculations

Since water is the dominant constituent in comets, the absorption of EUV photons in the coma is mainly dictated by $H_2O$. Therefore, we have considered a pure $H_2O$ comet at 1 AU with 4 different cases of gas production rate: (a) Q = $1\times10^{31}$ $s^{-1}$, (b) Q = $7\times10^{29}$ $s^{-1}$, (c) Q = $1\times10^{28}$ $s^{-1}$, and (d) Q = $1\times10^{32}$ $s^{-1}$. The first case is representative of C/1995 O1 Hale Bopp-type comets, the second case of 1P/Halley-type comets, the third case of 46P/Wirtanen-type comets, while the fourth case is taken to assess the effects of very high Q comets. The neutral density of $H_2O$ in the coma is calculated using Haser's model as described in our previous papers [*Bhardwaj et al.*, 1990, 1996; *Bhardwaj*, 1999], and the coma is assumed to be spherically symmetric. All the calculations are made at standard heliocentric distance of 1 AU for solar minimum conditions, which is quite appropriate for the Halley and Hale Bopp apparitions in 1986 and 1997, respectively. The solar EUV reference spectrum is taken from *Torr and Torr* [1979] in 41-1025 Å range. The $H_2O$ photoabsorption and photoionization cross sections are taken from *Haddad and Samson* [1986], which are binned into 71-wavelength intervals of *Torr and Torr* [1979] solar EUV flux.

The ion production rate, $q(r,\theta)$, at cometocentric distance r and solar zenith angle $\theta$ in an atmosphere due to the absorption of solar EUV radiation is given by

$$q(r,\theta) = \int_\lambda n(r)\, \sigma^I(\lambda)\, I_\infty(\lambda)\, \exp[-\tau(r,\theta)]\, d\lambda \qquad (1)$$

where
$$\tau(r,\theta) = \sigma^A(\lambda) \int n(r)\, ds \qquad (2)$$

Here $\sigma^A(\lambda)$ and $\sigma^I(\lambda)$ are the absorption and ionization cross sections at wavelength $\lambda$, $I_\infty(\lambda)$ is the unattenuated solar flux at top of the atmosphere at wavelength $\lambda$, $n(r)$ is the neutral density at cometocentric distance r, and ds is the element of distance along the path of radiation. Since our aim is to study the effects of SZA on photoionization rates we have used the generalized Chapman function [*Green and Martin*, 1966] to calculate optical depth. The use of generalized Chapman function is important as it allows for proper accounting of the optical depth effects at large SZA values.

For $\theta \leq 90°$ the optical depth $\tau = \sigma \int n\, ds$ is given by

$$\tau = \sigma \int_Z^\infty n(y)\, dy\, \sec\theta_y \qquad (3)$$

where
$$\sec\theta_y = 1\bigg/\left[1 - \left(\frac{R+Z}{R+y}\right)^2 \sin^2\theta\right]^{1/2} \tag{4}$$

Integration of the above equation gives

$$\tau = \sigma\, n(r)\, r\, (1/\sin\theta)\left[\pi/2 - \cos^{-1}\sin\theta\right] \tag{5}$$

where $r = Z + R$ is the cometocentric distance and R is the radius of the comet, which is taken as 10 km. For $\theta > 90°$ the optical depth $\tau$ is given by

$$\tau = \sigma\, n(r)\, r\, (1/\sin\theta)\left[\pi/2 + \cos^{-1}\sin\theta\right] \tag{6}$$

Ion production rate is obtained by carrying out integrations in equations (1) and (2).

## 3. Results and Discussion

Figure 1 shows the $H_2O^+$ photoion production rates for 4 different Q (gas production rate) values at SZA=0°. At a given radial distance, as Q increases the density of neutral $H_2O$ also increase resulting in a larger attenuation of solar radiation, thereby reducing the magnitude of peak production and upward shift of the ion production rate peak. However, at $Q=1\times10^{31}$ s$^{-1}$ a second peak in ion production rate profile is observed close to the nucleus. This lower peak is clearly seen at $Q=10^{32}$ s$^{-1}$. Also observed from Figure 1 is that the altitude region in the inner cometary coma over which the major solar EUV deposition takes place extend up to ~100 km from the nucleus at $Q=7\times10^{29}$ s$^{-1}$, to ~2000 km at $1\times10^{31}$ s$^{-1}$, and to ~$10^4$ km at $1\times10^{32}$ s$^{-1}$. Thus, in case of Halley-type comets the UV opacity effects are most important mainly within a few 100 km from the nucleus, which is consistent with the findings by earlier studies (e.g., *Giguere and Huebner*, 1978; *Huebner*, 1985, and references therein). The optical depth effects are generally neglected in data interpretation and modeling studies since most of the ion-modeling studies, which start basically with photoionization as the primary source, concentrate their efforts in doing comparison with the Giotto data that are available at radial distances >~$10^3$ km [e.g., *Balsiger et al.*, 1986; *Krankowsky et al.*, 1986]. But in case of bright comets, like Hale Bopp, opacity effects extend to a few $10^3$ km from the nucleus, and this distance increases with increase in Q, and hence cannot be ignored in modeling the inner cometary coma.

The reason for the occurrence of a double peak structure in photoion production profile seen in Figure 1 in higher Q comets can be understood on inspecting Figure 2, where the altitudinal degradation of solar flux at 3 selected wavelengths of 303.8 Å (40.8 eV), 180-165 Å (68.9-75.2 eV), and 62-41 Å (200-300.4 eV) are plotted at SZA=0° for 4 representative Q values. In case of $Q=1\times10^{28}$ s$^{-1}$ (low-Q comets), essentially no attenuation of solar EUV flux in the coma takes place, which is reflected in a straight-line profile of its photoionization rate seen in Figure 1. Even in case of brighter comets like Halley ($Q=7\times10^{29}$ s$^{-1}$), the neutral densities in the innermost coma are not sufficiently large to completely attenuate higher (>~60 eV) energy solar photons, while solar soft x-ray (>100 eV) photons travel almost unattenuated to the nucleus. The photoionization peak observed in Figure 1 at ~30-40 km is due to the attenuation of photons of energy <50 eV in the cometary atmosphere, with larger contribution from He II Lyman-$\alpha$ line at 303.8 Å, which is the dominant line in the solar EUV spectrum. As observed from Figure 2, the maximum attenuation of this line occurs in the 20-60 km region and hence the ion production peaks around these radial distances. At higher Q, the column density in the coma increases resulting in a larger attenuation even for higher energy photons. At $Q=1\times10^{32}$ s$^{-1}$, large attenuation of the He II line takes place ~4-10×10$^3$ km, while degradation of solar soft x-ray photons occurs in the innermost coma at ~50-100 km: giving rise to the well-defined double-peak structure in the photoionization rate profile.

Our calculated total photoionization rate of $H_2O$ at 1 AU is 4.4×10$^{-7}$ s$^{-1}$, which is in agreement, within about 10%, with values 4.1×10$^{-7}$ s$^{-1}$ and 3.8×10$^{-7}$ s$^{-1}$ reported by *Huebner et al.* [1992] and *Korosmezey et al.* [1987], respectively. The small differences are due to different solar EUV flux used in different studies, and partially due to photoionization cross sections used, which we have taken from *Haddad and Samson* [1986], while *Huebner et al.* [1992] compiled them from different sources.

In the post-Halley era, the electron impact ionization in the cometary coma has drawn the attention of several workers [e.g., *Boice et al.*, 1986; *Korosmezey et al.*, 1987; *Cravens et al.*, 1987; *Bhardwaj et al.*, 1990, 1996; *Wegmann et al.*, 1987, 1999; *Haider at al.*, 1993; *Haberli et al.*, 1996; *Bhardwaj*, 1999]. Evidences for the presence of an electron impact source in the cometary coma have been inferred from many observations made during the appearance of comet Halley, which are discussed in our earlier papers [e.g., *Bhardwaj et al.*, 1996; *Bhardwaj*, 1999]. The presence of electrons in cometary comae is clearly publicized by the detection of OI 1356 Å emission in comets [e.g., *McPhate et al.*, 1999], since this emission being a spin-forbidden transition cannot be produced by solar fluorescence.

The $H_2O^+$ ion production rates due to photoelectron impact ionization of $H_2O$ at 4 different Q values are also presented in Figure 1. The photoelectrons are generated in photoionization of $H_2O$ by solar photons whose wavelength is less than 984 Å, while at $\lambda<620$ Å the efficiency of ionization is unity (i.e., all solar photons absorbed by $H_2O$ in the coma create $H_2O^+$ ions and photoelectrons). The photoelectron impact ionization rates are calculated using Analytical Yield Spectrum (AYS) approach and inputs as described in our previous papers [e.g., *Bhardwaj et al.*, 1990, 1996; *Haider et al.*, 1993; *Bhardwaj*, 1999; *Bhardwaj and Haider*, 2002 and references therein]. The AYS is generated by running the Monte Carlo model and then analytically representing the yield [e.g., *Singhal and Green*, 1981; *Green et al.*, 1985; *Singhal and Bhardwaj*, 1991; *Bhardwaj and Michael*, 1999a,b], and is used to calculate the photoelectron flux. The photoelectron impact ion production rate, P(r), is calculated using

$$P(r) = n(r) \int_W F(E,r)\,\sigma(E)\,dE \qquad (7)$$

where W is the ionization threshold, $\sigma(E)$ is the ionization cross section at energy E, n(r) is the neutral gas density at cometcentric distance r, and F(E, r) is the photoelectron flux obtained by using the AYS technique (see *Bhardwaj et al.*, 1990, 1996; *Bhardwaj*, 1999 for details).

From Figure 1 it is seen that photoelectron impact $H_2O^+$ production rate is higher than photoionization rate at the peak. This is because longer wavelengths, which contribute to photoionization but do not contribute greatly to photoelectron flux, are attenuated more efficiently in the upper region of the atmosphere (due to larger absorption cross section at longer wavelengths) than the shorter wavelengths that actually produce energetic photoelectrons. However, in case of higher-Q comets, the photoelectron impact ionization rate at the second-lower photoionization peak is an order of magnitude larger than the photoionization rate. This can be understood from Figure 3 where we have plotted the energy spectrum of photoelectrons for the case $Q=10^{32}$ $s^{-1}$ at 4 different cometocentric distances of 60, 600, 6000, and 60,000 km. It can be noted that photoelectron energy spectrum at higher (>100 eV) energies is higher at shorter (60 and 600 km) distances compared to those at larger (>~6000 km) distances. This is because the absorption of solar soft x-rays occurs deep in the coma producing energetic (>100 eV) photoelectron which is capable of causing multiple ionization of $H_2O$ leading to a sharp increase in the ionization rate. But in case of $Q=7\times10^{29}$ $s^{-1}$, even at cometocentric distance of 30 km (photoionization peak), the >60-eV electrons produced are less compared to those produced at lower energies (cf. Figure 3). This is because the coma is not sufficiently dense to attenuate the solar soft x-ray flux, which is also evident from Figure 2.

The effect of large Q on the ion production rates is more emphatically demonstrated when we calculated the production rates at different SZA. In Figure 4 we present the photoionization rate profiles at SZA of 30°, 120°, and 160° at 4 Q values. It is seen that at $Q=7\times10^{29}$ $s^{-1}$ the peak production rate is obtained at around 40 km in the sunward direction and rises to an altitude ~1000 km in the anti-sunward direction. Also to be noticed is the fact that the second peak structure starts to appear when SZA>130° in case of comets with lower Q of $7\times10^{29}$ $s^{-1}$, since photons now travel a larger column resulting in an increased attenuation. At $Q=1\times10^{31}$ $s^{-1}$ the primary (upper) peak rises from ~500 km in sunward to around $10^4$ km in anti-sunward direction, and the production rate at peak falls by 2 orders of magnitude, from ~100 $cm^{-3}$ $s^{-1}$ to ~1 $cm^{-3}$ $s^{-1}$. Moreover, the lower peak is roughly of the same magnitude as the upper peak. In case of $Q=1\times10^{32}$ $s^{-1}$, the opacity effects are found to extend to ~$10^5$ km. This has implications for deriving quantities from observations, like production rate from column brightness, particularly in higher Q-comets, because one cannot assume the coma to be uniformly sunlit: the absorption effects on the solar radiation has to be taken into account. Such sunward to anti-sunward differences may be one of the causes of asymmetry seen in the emission profiles of metastable atomic O and C emissions on comet Hale-Bopp [*Morgenthaler et al.*, 2002; *Oliversen et al.*, 2002]. The SZA effects will be more pronounced when the field-of-view of the instrument is relatively smaller [e.g., *Oliversen et al.*, 2002].

The double peak structure in the $H_2O^+$ ion production rate profile predicted by the present calculations in the inner coma of large-Q comets can be observed either by in-situ measurements, like the Giotto mission for comet Halley, or by measuring emissions induced by electrons at spatial resolution sufficient to resolve the structure, like OI 1356 Å emission. Alternately, the double peak structure can be detected by observing $H_2O^+$ ion emissions at sufficiently high spatial resolution of a few 100 km, although in this case a model [e.g., *Wegmann et al.*, 1999] would be required to do a better interpretation, because of the chemistry involved. It may be mentioned here that in high-Q comets the size of the nucleus could be larger (say ~50 km at $Q=10^{32}$ s$^{-1}$) than the 10-km radius assumed for all Q values in the present study.

## 4. Summary

In this paper we have emphasized the need for a proper accounting of the absorption of solar EUV radiation in the inner coma of comets with large gas production rates. In higher-Q comets, a double-peak structure in photoionization rate profile is found, which results from the effective degradation of solar soft x-ray photons deep in the coma: producing energetic photoelectrons. These energetic photoelectrons create additional ionization in the coma whose magnitude at the peak is more than a factor of 10 larger than the photoionization peak. The effect of SZA on the photoionization rates is studied and is found to be important in higher-Q comets. The present study has demonstrated the importance of photoelectron impact as the ionization source relative to solar EUV radiation in the inner coma of high-Q comets.

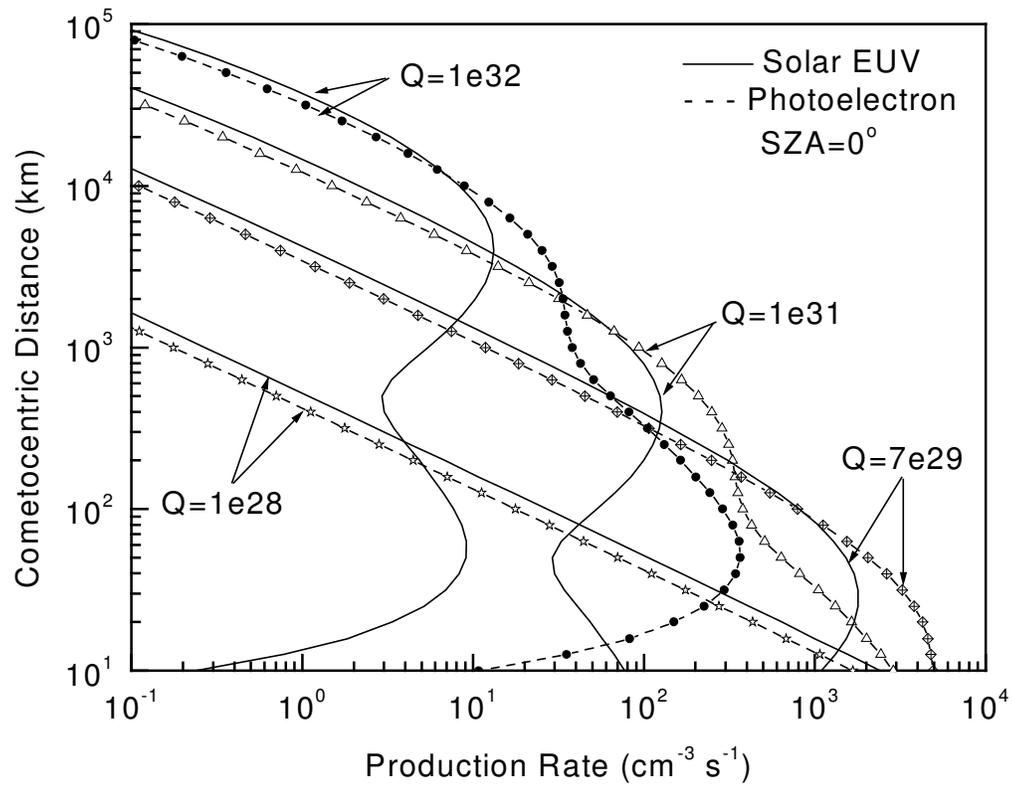

**Figure 1.** Photoionization and photoelectron impact ionization rates of $H_2O^+$ at solar zenith angle of $0°$ for 4 different gas production rates Q (in $s^{-1}$).

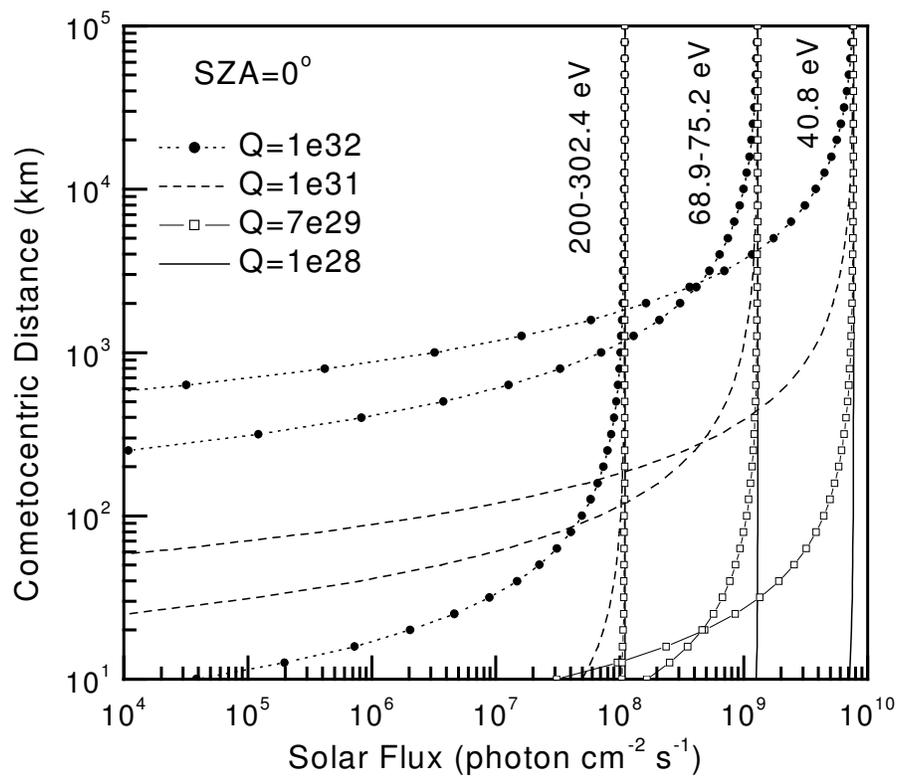

**Figure 2.** Radial degradation profiles of solar EUV flux at 3 energies at solar zenith angle of 0°  for 4 different gas production rates Q (in s$^{-1}$).

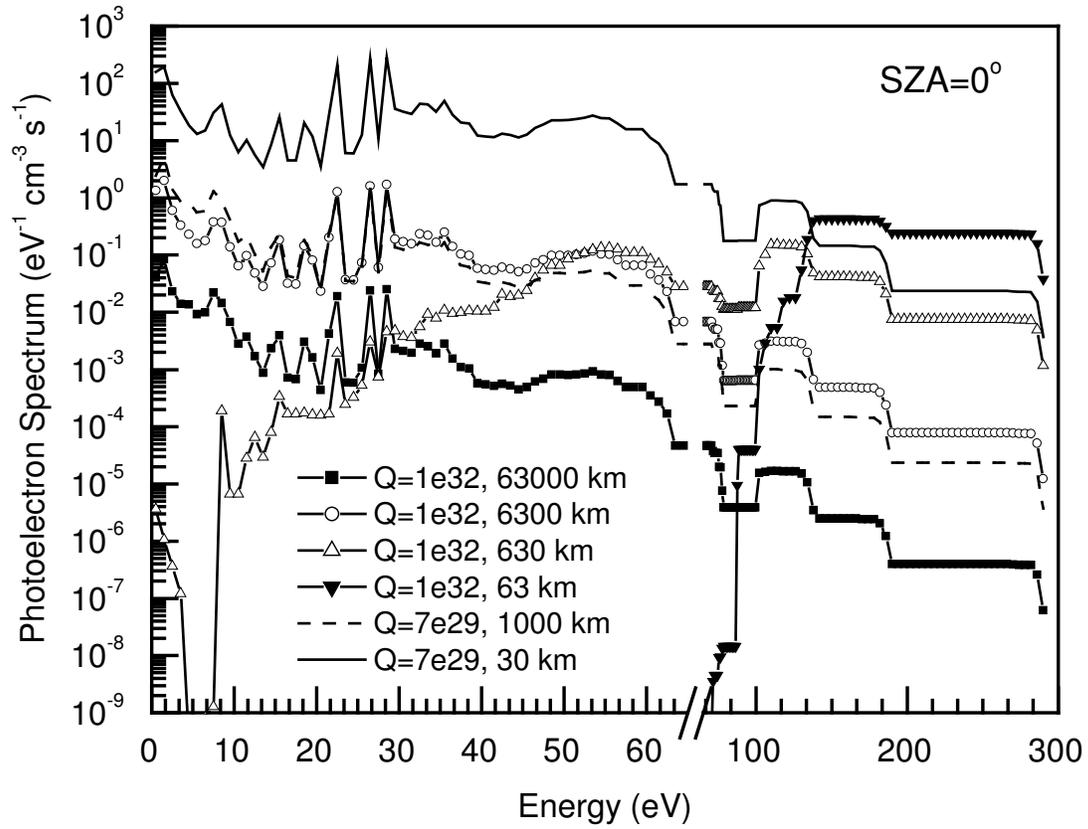

**Figure 3.** Photoelectron energy spectrum at 4 cometocentric distances in case of Q=1×10$^{32}$ s$^{-1}$, and at 2 distances in case of Q=7×10$^{29}$ s$^{-1}$. Note the change in the energy scale on x-axis after break at 70 eV.

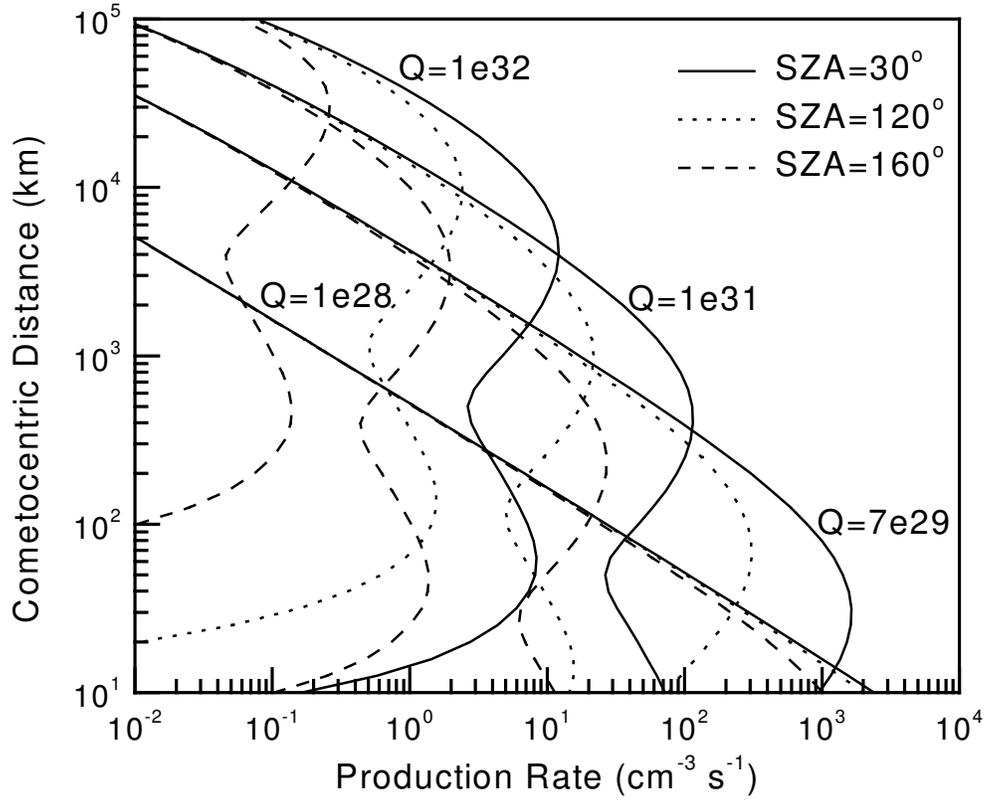

**Figure 4.** Production rate of $H_2O^+$ ion due to photoionization of $H_2O$ for 4 different gas production rates Q (in $s^{-1}$), at 3 solar zenith angles of $30°$, $120°$, and $160°$.